\documentclass[12pt]{iopart}

\usepackage{iopams}
\begin{document}

\title{Static Dipole Polarizability for the $1s\sigma$ electronic
state of the $\mathrm{H^{+}_{2}}$ molecular ion}

\author{Ts.~Tsogbayar}

\address{Institute of Physics and Technology, Mongolian Academy of Sciences,
Peace Avenue 54B, 210651, Ulaanbaatar 51, Mongolia}
\ead{tsogbayar@ipt.ac.mn}
\begin{abstract}
The static dipole polarizabibility for the $1s\sigma$ electron
state of the $\mathrm{H}_{2}^{+}$ hydrogen molecular ion is
calculated within Born-Oppenheimer approximation. The variational
expansion with randomly chosen exponents has been used for
numerical studies. The results obtained for the dipole
polarizability are accurate to the nine digits.

\end{abstract}

\maketitle

\section{Introduction}
One of sample few-body systems, a $\mbox{H}^+_2$ molecular ion
have played vital roles in the development of molecular quantum
mechanics. Moreover, this ion has been considered as a model
system for the formulation of many different methods and
approximations and is much easer to study theoretically than
experimentally, and although its physical properties have been
extensively calculated, few have been measured so far.

At present, the system has been redeemed widely not only in
theoretical field but in experimental sites. New type of
variational expansions based on randomly chosen exponents
\cite{kor1,bai} have been employed successfully by many
researchers for the calculation of non-relativistic energy for the
vibrational $v=0$ and rotational $J=0$ electronic state, even more
the relativistic and radiative corrections of higher order of
$m\alpha$ for $1S\sigma (v=0,J=0)$ state of the $\mbox{H}^{+}_{2}$
molecular ion \cite{mark,kor6}. Within the Born-Oppenheimer
approximation, the improved relativistic corrections of
$m\alpha^{4}$ and $m\alpha^{6}$ orders for the ground state
$(v=0,J=0)$ of the $\mbox{H}^{+}_{2}$ molecular ion have been
obtained in \cite{kor5,tsog}.

There are, in addition, many theoretical precision nonadiabatic
values of the electric-dipole polarizability of the ground state
$(v=0,J=0)$ of the $\mbox{H}^+_2$ molecular ion. Shertzer and
Greene \cite{sher} had evaluated the polarizabilities of the
ground state $(v=0,J=0)$ of the $\mbox{H}^{+}_{2}$ and
$\mbox{D}^{+}_{2}$ using the finite element method. Bhatia and
Drachman \cite{bhat} calculated the dipole polarizability of the
ground states $(v=0,J=0)$ of the ions $\mbox{H}^{+}_{2}$ and
$\mbox{D}^{+}_{2}$ using the second-order perturbation theory in
Hylleraas coordinate. Taylor \textit{et al} \cite{tayl} obtained
the electric-dipole polarizabilities of the ground state of
$\mbox{H}^{+}_{2}$ and $\mbox{D}^{+}_{2}$ variationally using
traditional molecular physics method in which any approximation
based on the size of the electron mass relative to the nucleus
mass could be avoided. Moss \cite{moss} obtained rotationally
averaged polarizabilities of the $\mbox{H}^{+}_{2}(0,0)$ and
$\mbox{D}^{+}_{2}(0,0)$ from scattering theory. Yan \textit{et al}
\cite{zong} had derived more accurate theoretical values of the
both nonrelativistic energy and dipole polarizability of ground
state $(v=0,J=0)$ of $\mbox{H}^{+}_{2}$ using variational wave
function in Hylleraas coordinates. Hilico \textit{et al}
\cite{hil1} had derived the most accurate results for the ground
states $(v=0,J=0)$ of hydrogen molecular ions $\mbox{H}^{+}_{2}$
and $\mbox{D}^{+}_{2}$ using the perimetric coordinates. Korobov
\cite{kor4} had calculated the lowest order relativistic
correction to the electric-dipole polarizability of the ground
state of the $\mbox{H}^{+}_{2}$ molecular ion.

Bishop and co-workers \cite{bish1,bish2,bish3} had obtained the
dipole polarizabilities of the ground state $(v=0,J=0)$ of the
ions in a clamped nucleus approximation. Various approaches for
obtaining the polarizability in the adiabatic approximation were
considered in \cite{bish4}.

Static dipole polarizability of the $1S\sigma (v=0,J=0)$
electronic state of the the $\mbox{H}^{+}_{2}$ molecular ion had
been calculated theoretically many times in the Born-Oppeheimer
approximation \cite{mcEac,rah,bates,cal,ad,dal}.

Furthermore, several laser spectroscopy experiments have been
proposed \cite{gr,schil,roth} for high precision measurements of
the rovibrational spectrum of hydrogen molecular ions
$\mbox{H}^{+}_{2}$ and $\mbox{HD}^{+}$.

The precision measurement of the scalar electric-dipole
polarizability for the $1S\sigma (v=0,J=0)$ state of the
$\mbox{H}^+_2$ molecular ion have been performed by Jacobson
\textit{at al} \cite{jac1,jac2}.

In this work our goal is to obtain more accurate values for the
static  dipole polarizability for the $1S\sigma (v=0,J=0)$ state
of $\mbox{H}^+_2$ than in \cite{bish1} in the Born-Oppenheimer
approximation.

We want to show that the use of the variational expansion
suggested in this work allows an analytical evaluation of the
singular matrix elements required for the relativistic
calculations and can provide us with very accurate data.

In future studies it can be used to obtain the lowest order
relativistic correction to the mean electronic dipole
polarizability of the $1S\sigma (v=0,J=0)$ state of $\mbox{H}^+_2$
in the Born-Oppenheimer approximation.

In Sect.\,2, we outline the theoretical method for obtaining the
the static dipole polarizability for the $1S\sigma (v=0,J=0)$
state of $\mbox{H}^+_2$ in the Born-Oppenheimer approximation. In
Sect.\,3 we consider the variational wave function and the
nonrelativistic energy for this state of the $\mbox{H}^{+}_{2}$.
Moreover, main results of numerical calculation will be presented
in Tables in this section. In Sect.\,4 we compare our results with
the other values obtained in the Born-Oppenheimer approximation.
Then a conclusion will be discussed.

\section{Theory}

The Schr\"{o}dinger equation for the $\mbox{H}^{+}_{2}$ molecular
ion may be solved at various different levels of approximation. In
what follows, we will consider one of them; the Born-Oppenheimer
approximation which means that an electron moves in the field of
the clamped nuclei.

In the clamped nucleus approximation, the electronic motion is
perturbed by the electric field $\mathcal{E}$, and the mean
electronic dipole polarizability $\alpha (R)$ is derived as a
function of the internuclear separation $R$. In this work the
factor of $(1+\epsilon)=(2m+2)/(2m+1)$ (details can be found in
\cite{sher,bhat,tayl,moss,zong,hil1,kor4}, $m$ is a mass of
nucleus) in the perturbation will not be considered. Then
$\alpha(R)$ is averaged over the rovibronic state. In the sum over
rovibronic states approach, the adiabatic molecular wave function
is written as a product of an electronic and nuclear function. In
this paper, however, we do not consider the effects due to the
rovibronic state, and think over only the mean electronic
polarizability $\alpha (R)$ for ground state $v=0,J=0$ of
$\mbox{H}^{+}_{2}$ molecular ion.

In the Born-Oppenheimer approximation the electronic wave function
$\Psi_{0}(\mathbf{r};R)$ satisfies the following Schr\"{o}dinger
equation
\begin{equation}\label{eleq}
 \Big[-\frac{\hbar^{2}}{2 m_{e}}\nabla^{2}_{r} -\frac{Z_1}{r_1}-\frac{Z_2}{r_2}
 \Big]\Psi_{0}(\mathbf{r};R) = E_{0}(R) \Psi_{0}(\mathbf{r};R)\,.
\end{equation}
where $r_1$ and $r_2$ are the distances from an electron to nuclei
1 and 2, respectively.

Interaction with external electric field is given by
\begin{equation}\label{int}
H'= \mathbf{d}\cdot \mathbf{\mathcal{E}}
\end{equation}
where $\mathbf{d}$ is the dipole moment and $\mathbf{\mathcal{E}}$ is electric field strength.

Using the perturbation theory, it is easy to present that
\begin{equation}\label{enerex}
 E = E_{0} + E'' \mathbf{\mathcal{E}}^{2}
\end{equation}
to the second order in $\mathbf{\mathcal{E}}$, and $E_{0}$ is the
solution of Eq.(\ref{eleq}).

If the electric field is parallel to the nuclear axis, the polarizability is expressed by
\begin{equation}\label{dpar}
 \alpha_{\parallel} = -2 E'' 
\end{equation}
and if it is perpendicular to the nuclear axis, the polarizability is expressed by
\begin{equation}\label{dper}
 \alpha_{\perp} = -2 E'' 
\end{equation}
The mean electronic polarizability is calculated by
\begin{equation}\label{dmean}
 \alpha = \frac{1}{3}(\alpha_{\parallel} + 2 \alpha_{\perp}).
\end{equation}
It can be shown \cite{bish1,kolos} that
\begin{equation}\label{enesec}
 E''= \mathbf{B}^{\dagger} \mathbf{D} \mathbf{A}
\end{equation}
where $\mathbf{B}$ is the vector of coefficients in the expansion
of the first-order wave function
\begin{equation}\label{wff}
 \Psi'=\sum_{n}b_{n}\psi'_{n}
\end{equation}
($\psi_{n}$ is certain basic functions), $\mathbf{A}$ is the
vector of coefficients in the expansion of the zero-order wave
function
\begin{equation}\label{wfz}
 \Psi_{0}=\sum_{n}a_{n}\psi_{n}
\end{equation}
and $\mathbf{D}$ is the matrix elements with elements
\begin{equation}\label{dmat}
 D_{nm}=\langle \psi'_{n} d \psi_{m}\rangle \,.
\end{equation}
The $\mathbf{B}$ is obtained from the set of inhomogeneous
equations:
\begin{equation}\label{ineq}
 (\mathbf{H}-E_{0}\mathbf{S})\mathbf{B}=-\mathbf{D}\mathbf{A}
\end{equation}
where the matrix elements of $\mathbf{H}$ and $\mathbf{S}$ are
$\langle \psi'_{n} H_{0}\psi'_{m}\rangle$ and $\langle \psi'_{n}
\psi'_{m}\rangle$, respectively.

\section{Variational approximation and numerical results}

The variational wave function for the $1s\sigma (v=0,J=0)$ state
of $\mbox{H}^+_2$ should be symmetrized and is constructed as
follows
\begin{equation}\label{wf}
   \Psi_{0} = \sum^{\infty}_{i=1}
 a_{i}(e^{-\alpha_{i} r_{1} - \beta_{i} r_{2}} +
       e^{-\beta_{i} r_{1} - \alpha_{i} r_{2}} ).
\end{equation}
Parameters $\alpha_{i}$ and $\beta_{i}$ are generated in a
quasi-random manner
\begin{equation}\label{par}
  \alpha_{i}=\lfloor\frac{1}{2}i(i+1)\sqrt{p_{\alpha}}\rfloor(A_{2}-A_{1})
  + A_{1}
\end{equation}
$\lfloor x\rfloor$ designates the fractional part of $x$,
$p_{\alpha}$ is a prime number, an interval $[A_{1},A_{2}]$ is a
real variational interval, which has to be optimized. Parameters
$\beta_{i}$ are obtained in a similar way. Details of the method
and discussion of various aspects of its application can be found
in \cite{kor1,bai}.

If the electric field is parallel to the nuclear axis, then
$$
d_{z}=-\frac{r^{2}_{1}-r^{2}_{2}}{2R} \,,
$$
and the perturbed function $\Psi'$ have a form
\begin{equation}\label{wfpar}
   \Psi'_{\parallel} = \sum^{\infty}_{i=1}
 b_{i}(e^{-\alpha_{i} r_{1} - \beta_{i} r_{2}} +
       e^{-\beta_{i} r_{1} - \alpha_{i} r_{2}} ).
\end{equation}
If the electric field is perpendicular to the nuclear axis, then
$$
 d_{x}=-r \cos\phi \,,
$$
and the perturbed function is taken the form
\begin{equation}\label{wfper}
   \Psi'_{\perp} = r\cos\phi \sum^{\infty}_{i=1}
 b_{i}(e^{-\alpha_{i} r_{1} - \beta_{i} r_{2}} +
       e^{-\beta_{i} r_{1} - \alpha_{i} r_{2}} ) \, ,
\end{equation}
where $r$ is a distance from center along internuclear axis,
$$
 r = \frac{1}{2 R}\sqrt{2
 r^{2}_{1}r^{2}_{2}+2r^{2}_{1}R^{2}+2r^{2}_{2}R^{2}-r^{4}_{1}-r^{4}_{2}-R^{4}}
 \,.
$$
In Table 1 we show variational parameters employed in the
calculation of energy values $E_{0}$ and unperturbed wave function
$\Psi_{0}$ at internuclear distance $R=2.0a.u.$. Using this type
of variational parameters leads to the very fast convergence.
\begin{table}[h]
\caption{Variational parameters and number of basic functions
$(n_{i})$ for different subsets of the variational wave function
with $N=100$. Intervals $[A_{1}, A_{2}]$ and $[B_{1}, B_{2}]$
correspond to part of a randomly chosen parameters $\alpha_{i}$
and $\beta_{i}$ [see Eq.(\ref{par}) for detials], respectively,
for the bond length $R=2.0\,a.u$. Prime numbers are $p_{\alpha}=2,
p_{\beta}=3$.}
\begin{center}
\begin{tabular}{c@{\hspace{3mm}}c@{\hspace{4mm}}c@{\hspace{4mm}}c@{\hspace{4mm}}c@{\hspace{4mm}}c@{\hspace{4mm}}}
\hline\hline
 & $N$ & $A_{1}$  & $A_{2}$ & $B_{1}$ & $B_{2}$ \\
\hline
i=1 & 50 & 0.00 & 1.50 & 0.00 & 0.40\\
i=2 & 50 & 0.20 & 2.00 & 1.00 & 6.00\\
\hline\hline
\end{tabular}
\end{center}
\end{table}

In Table 2 we present a comparison of the wave functions for the
ground state $(v=0,J=0)$ of the $\mbox{H}^{+}_{2}$ molecular ion
at bond length $R=2.0\,a.u.$. The variationally obtained values of
the $\Psi_{0}$ wave function are more accurate than the previous
ones.
\begin{table}[h]
\caption{The wave function for the $1s\sigma_{g}$ electron state
of the $\mathrm{H^{+}_{2}}$ molecular ion at bond length
$R=2.0\,a.u$.}
\begin{center}
\begin{tabular}{c@{\hspace{3mm}}c@{\hspace{4mm}}c@{\hspace{4mm}}c@{\hspace{4mm}}c@{\hspace{4mm}}c@{\hspace{4mm}}}
\hline\hline
& \multicolumn{1}{c}{Bates \textit{et al}, \cite{bates2}} & & & \multicolumn{1}{c}{Bates \textit{et al}, \cite{bates2}}  \\
\cline{2-2} \cline{5-5}
$r $ &       &     $\Psi_{0}(a.u.)$  & $r $  &  & $\Psi_{0}(a.u.)$\\
\hline
0.00 & 0.315 & 0.314\,692\,27 & 1.60 &       & 0.206\,542\,47\\
0.10 &       & 0.315\,970\,96 & 1.70 &       & 0.180\,471\,77\\
0.20 &       & 0.319\,824\,69 & 1.80 &       & 0.157\,611\,43\\
0.30 &       & 0.326\,306\,69 & 1.90 &       & 0.137\,581\,69\\
0.40 &       & 0.335\,506\,65 & 2.00 & 0.120 & 0.120\,044\,55\\
0.50 &       & 0.347\,552\,13 & 2.10 &       & 0.104\,699\,83\\
0.60 &       & 0.362\,610\,66 & 2.20 &       & 9.128\,154\,54(-2)\\
0.70 &       & 0.380\,892\,46 & 2.30 &       & 7.955\,434\,13(-2)\\
0.80 &       & 0.402\,653\,81 & 2.40 &       & 6.931\,038\,06(-2)\\
0.90 &       & 0.428\,201\,25 & 2.50 &       & 6.036\,634\,71(-2)\\
1.00 & 0.458 & 0.457\,896\,57 & 2.60 &       & 5.256\,076\,55(-2)\\
1.10 &       & 0.401\,762\,60 & 2.70 &       & 4.575\,156\,24(-2)\\
1.20 &       & 0.352\,207\,50 & 2.80 &       & 3.981\,386\,43(-2)\\
1.30 &       & 0.308\,524\,10 & 2.90 &       & 3.463\,801\,93(-2)\\
1.40 &       & 0.270\,066\,59 & 3.00 & 0.003 & 3.012\,782\,38(-2)\\
1.50 &       & 0.236\,248\,99 & 4.00 & 0.007 & 7.385\,569\,68(-3)\\
\hline\hline
\end{tabular}
\end{center}
\end{table}

In Table 3, the convergence of both $E_{0}(R=2.0a.u)$ and
$\Psi_{0}(r=2.0a.u;R=2.0a.u)$ is presented. In this calculation
the variational parameters presented in Table 1 are used.
\begin{table}[h]
\caption{Convergence of both energy at bond length $R=2.0\,a.u$
and wave function at distance from center along internuclear axis
$r=2.0\,a.u$ for the $1s\sigma_{g}$ electron state of the
$\mathrm{H^{+}_{2}}$ molecular ion.}
\begin{center}
\begin{tabular}{c@{\hspace{3mm}}c@{\hspace{4mm}}c@{\hspace{4mm}}}
\hline\hline
$N $ & $E_{0} (a.u.)$  & $\Psi_{0}(a.u.)$   \\
\hline
60 & \textbf{-1.102\,634\,214\,494\,946\,4}58\,49  & \textbf{0.120\,044\,550}\,742\,2\\
70 & \textbf{-1.102\,634\,214\,494\,946\,460\,6}0  & \textbf{0.120\,044\,550}\,805\,9\\
90 &\textbf{-1.102\,634\,214\,494\,946\,461\,50}   & \textbf{0.120\,044\,550}\,828\,1\\
100 & \textbf{-1.102\,634\,214\,494\,946\,461\,50} & \textbf{0.120\,044\,550}\,828\,3 \\
\hline\hline
\end{tabular}
\end{center}
\end{table}

\begin{table*}[t]
\caption{The dipole polarizability for the $1s\sigma (v=0,J=0)$
 state of $\mathrm{H}_{2}^{+}$}
\begin{center}
\begin{tabular}{c@{\hspace{3mm}}c@{\hspace{4mm}}c@{\hspace{4mm}}
   c@{\hspace{4mm}}c@{\hspace{4mm}}c}
\hline\hline
& \multicolumn{2}{c}{Bishop, Cheung \cite{bish1}}  \\
\cline{2-3}
$R$ & $\alpha_{\parallel}$  & $\alpha_{\perp}$ & $\alpha_{\parallel}$  & $\alpha_{\perp}$ & $\alpha$ \\
\hline 0.05 &  &  &  0.284\,434\,63 &
   0.284\,064\,67 & 0.284\,187\,99 \\
0.10 &  &  & 0.293\,065\,94 &
   0.291\,556\,44 & 0.292\,059\,61 \\
0.20 & 0.323\,181 & 0.316\,745 & 0.323\,180\,58 &
   0.316\,745\,01 & 0.318\,890\,20 \\
0.30 &  &  & 0.367\,729\,08 &
   0.352\,002\,47 & 0.357\,244\,67 \\
0.40 & 0.425\,732 & 0.395\,077 & 0.425\,732\,10 &
   0.395\,076\,94 & 0.405\,295\,33 \\
0.50 &  &  & 0.497\,558\,14 &
   0.444\,803\,03 & 0.462\,388\,07 \\
0.60 & 0.584\,328 & 0.500\,513 & 0.584\,328\,47 &
   0.500\,513\,40 & 0.528\,451\,76 \\
0.70 &  &  & 0.687\,676\,16 &
   0.561\,784\,00 & 0.603\,748\,05 \\
0.80 & 0.809\,650 & 0.628\,314 & 0.809\,649\,71 &
   0.628\,314\,08 & 0.688\,759\,29 \\
0.90 &  &  & 0.952\,682\,22 &
   0.699\,865\,38 & 0.784\,137\,66 \\
1.00 & 0.111\,959(1) & 0.776\,229 & 0.111\,959\,40(1) &
   0.776\,229\,40 & 0.890\,684\,27 \\
1.30 &  &  & 0.179\,816\,83(1) &
   0.103\,221\,75(1) & 0.128\,753\,44(1) \\
1.50 &  &  & 0.244\,196\,01(1)&
   0.122\,321\,10(1) & 0.162\,946\,07(1) \\
1.80 & 0.380\,895(1) & 0.153\,548(1) & 0.380\,894\,55(1) &
   0.153\,548\,15(1) & 0.229\,330\,28(1) \\
2.00 & 0.507\,765(1) & 0.175\,765(1) & 0.507\,764\,90(1) &
   0.175\,764\,86(1) & 0.286\,431\,54(1) \\
2.30 &  &  & 0.772\,496\,21(1) &
   0.210\,563\,48(1) & 0.397\,874\,39(1) \\
2.50 &  &  & 0.101\,494\,85(2) &
   0.234\,342\,47(1) & 0.494\,544\,48(1) \\
2.80 & 0.151\,542(2) & 0.270\,145(1) & 0.151\,541\,85(2) &
   0.270\,144\,97(1) & 0.685\,236\,15(1) \\
3.00 & 0.196\,995(2) & 0.293\,661(1) & 0.196\,995\,06(2) &
   0.293\,660\,88(1) & 0.852\,424\,12(1) \\
3.50 &  &  & 0.374\,330\,77(2) &
   0.349\,066\,59(1) & 0.148\,048\,03(2) \\
4.00 & 0.700\,473(2) & 0.396\,437(1) & 0.700\,473\,12(2) &
   0.396\,437\,11(1) & 0.259\,920\,18(2) \\
4.50 &  &  & 0.129\,572\,67(3) &
   0.433\,014\,66(1) & 0.460\,776\,54(2) \\
5.00 & 0.237\,462(3) & 0.457\,944(1) & 0.237\,462\,05(3) &
   0.457\,943\,66(1) & 0.822\,069\,74(2) \\
5.50 &  &  & 0.431\,696\,21(3) &
   0.472\,234\,01(1) & 0.147\,046\,96(3) \\
6.00 & 0.779\,110(3) & 0.478\,162(1) & 0.779\,109\,89(3) &
   0.478\,162\,22(1) & 0.262\,891\,04(3) \\
6.50 &  &  & 0.139\,675\,65(4) &
   0.478\,432\,00(1) & 0.468\,775\,05(3) \\
7.00 & 0.248\,888(4) & 0.475\,481(1) & 0.248\,888\,50(4) &
   0.475\,481\,44(1) & 0.832\,798\,21(3) \\
7.50 &  &  & 0.441\,080\,53(4) &
   0.471\,132\,49(1) & 0.147\,340\,93(4) \\
8.00 & 0.777\,895(4) & 0.466\,548(1) & 0.777\,895\,05(4) &
   0.466\,548\,44(1) & 0.259\,609\,38(4) \\
8.50 &  &  & 0.136\,601\,77(5) &
   0.462\,359\,95(1) & 0.455\,647\,47(4) \\
9.00 & 0.238\,968(5) & 0.458\,835(1) & 0.238\,967\,73(5) &
   0.458\,835\,24(1) & 0.796\,864\,99(4) \\
9.50 &  &  & 0.416\,635\,75(5) &
   0.456\,027\,67(1) & 0.138\,908\,99(5) \\
10.00 & 0.724\,216(5) & 0.453\,880(1) & 0.724\,215\,81(5) &
   0.453\,880\,28(1) & 0.241\,435\,53(5) \\
\hline\hline
\end{tabular}
\end{center}
\end{table*}

\begin{table}[h]
\caption{Convergence of dipole polarizabilities with the increase
in the basic-set size of $\Psi'$}
\begin{center}
\begin{tabular}{c@{\hspace{3mm}}c@{\hspace{4mm}}c@{\hspace{4mm}}c@{\hspace{4mm}}}
\hline\hline
$R (a.u.)$ & $N $ & $\alpha_{\parallel}$  & $\alpha_{\perp}$   \\
\hline
1.00 & 100 & \textbf{0.111\,959\,39}1\,834 (1) & \textbf{0.776\,229\,39}4\,339\\
     & 150 & \textbf{0.111\,959\,395}\,159 (1) & \textbf{0.776\,229\,397}\,334\\
     & 200 & \textbf{0.111\,959\,395\,159} (1) & \textbf{0.776\,229\,397\,3}34\\
     & 250 & \textbf{0.111\,959\,395\,159} (1) & \textbf{0.776\,229\,397\,33}5\\
     & 300 & \textbf{0.111\,959\,395\,159} (1) & \textbf{0.776\,229\,397\,335}\\
1.50 & 100 & \textbf{0.244\,19}5\,787\,807 (1) & \textbf{0.122\,321\,0}61\,209(1)\\
     & 150 & \textbf{0.244\,196\,012\,3}82 (1) & \textbf{0.122\,321\,09}9\,085(1)\\
     & 200 & \textbf{0.244\,196\,012\,392} (1) & \textbf{0.122\,321\,099}\,095(1)\\
     & 250 & \textbf{0.244\,196\,012\,392} (1) & \textbf{0.122\,321\,099\,09}5(1)\\
     & 300 & \textbf{0.244\,196\,012\,392} (1) & \textbf{0.122\,321\,099\,095}(1)\\
2.00 & 100 & \textbf{0.507\,7}59\,939\,512 (1) & \textbf{0.175\,76}3\,635\,406(1)\\
     & 150 & \textbf{0.507\,764\,89}6\,289 (1) & \textbf{0.175\,764\,}860\,893(1)\\
     & 200 & \textbf{0.507\,764\,898\,343} (1) & \textbf{0.175\,764\,86}1\,787(1)\\
     & 250 & \textbf{0.507\,764\,898\,353} (1) & \textbf{0.175\,764\,861\,7}93(1)\\
     & 300 & \textbf{0.507\,764\,898\,353} (1) & \textbf{0.175\,764\,861\,793}(1)\\
\hline\hline
\end{tabular}
\end{center}
\end{table}

In Tables 4 the mean electronic polarizability for the $1s\sigma$
electronic state of $\mbox{H}^{+}_{2}$ molecular ion and
comparison with \cite{bish1} are presented. The estimated accuracy
of obtained values is nine significant digits. In Table 5 the
convergence of $\alpha_{\parallel}$ and $\alpha_{\perp}$ with the
increase in the expansion length of $\Psi'_{\parallel}$ and
$\Psi'_{\perp}$ (Eqs.(\ref{wfpar})-(\ref{wfper})) is demonstrated
for some values of internuclear distance $R$.

Numerical evaluation of the matrix elements for operators in
(\ref{dpar}) and (\ref{dper}) is expounded in the Appendix.

In order to get accurate results we use three sets of basic
function of the type (\ref{wf}) (in a spirit of \cite{kor1}) for
small values of internuclear distance $R$, two sets for
intermediate and large values of $R$, respectively. Total number
of the basic function varies from $N=110$ to $N=300$. In our
calculations arithmetics of sextuple precision (about 48 decimal
igits) implemented as a FORTRAN90 module has been used. In all
tables the factor $x$ in the brackets means $10^{x}$. Atomic units
are used throughout.

\section{Conclusion}

The mean electronic polarizabibility for the $1s\sigma (v=0,J=0)$
electron state of the $\mathrm{H}_{2}^{+}$ hydrogen molecular ion
have been accurately calculated within Born-Oppenheimer
approximation. The variational expansion with randomly chosen
exponents has been used for numerical studies. This type of
expansion allows us to use few number of basic functions. If
nonrelativistic energy values of $\mbox{H}^{+}_{2}(0,0)$ are
required to $10^{-15}\,a.u.$ accuracy, then the wave functions
must be accurate, at least, to this same level. The wave
functions, however, are typically accurate to less than half as
many significant figures as the energy. Seeing from Tables 3, the
precision nonrelativistic energy value $E_{0}$ at bond length $
R=2.0\,a.u$ can easily reach to 15-significant digits for basic
function number $N=60$ and the unperturbed wave function
$\Psi_{0}$ have 10-significant digits using the same number of
terms.

Previous calculations performed for $1S\sigma (v=0,J=0)$ state of
$\mbox{H}^{+}_{2}$ molecular ion over a wide range of internuclear
separations $R$ are by Dalgarno and Lewis \cite{dal}, Calvert and
Davison \cite{cal}, Bates \cite{bates}, McEachran and Smith
\cite{mcEac}, and Bishop and Cheung \cite{bish1}. Both Dalgarno
and Lewis \cite{dal} and Bates \cite{bates} used the oscillator
strength sum rule. McEachran and Smith \cite{mcEac} employed the
variational procedure, using the accurate two-center James
\cite{jam} orbital as the perturbed function. Bishop and Cheung
\cite{bish1} had calculated variationally the first accurate mean
electronic polarizability over a wide range of internuclear
separations. Rahman \cite{rah} and Adamov \textit{et al} \cite{ad}
had performed accurate first order variational calculations near
the equilibrium distance $R=2.0\,a.u.$.

\begin{table}
\caption{Comparison with earlier calculations at a bond length
$R=2.0\,a.u.$}
\begin{center}
\begin{tabular}{@{}c@{\hspace{6mm}}l@{\hspace{6mm}}l@{\hspace{6mm}}l@{}}
\hline\hline
 & \multicolumn{1}{c}{$\alpha_{\parallel}$~~~~} & \multicolumn{1}{c}{$\alpha_{\perp}$~~~~}
 \\
\hline \cite{rah}
& $5.061$ & $1.758$  \\
\cite{dal}
& $5.283$ & $2.202$ \\
\cite{ad}
& $5.173$ & $1.847$ \\
\cite{cal}
& $5.084$ & $1.767$ \\
\cite{bates}
& $5.06$ & $1.76$ \\
\cite{mcEac}
& $5.199$ & $1.829$  \\
\cite{bish1}
& $5.077\,65$ & $1.757\,65$  \\
this work
& $5.077\,6490$ & $1.757\,6486$ \\
\hline\hline
\end{tabular}
\end{center}
\end{table}

In Table 6 we place the comparison of our results with the earlier
ones, which demonstrate superiority of the newly obtained results.
The results obtained for the dipole polarizability are accurate to
the nine digits. That is two digits more accurate than in
\cite{bish1}.

In general, the total molecular rovibronic wave function is taken
a form as a product of separate electronic, vibrational and
rotational wave functions, that is, $\Psi_{vJM}=|v(J)\rangle
Y^{M}_{J}(\theta , \phi)$--rovibrational wave function. Then the
full static dipole polarizability can be written in the form
\begin{equation}\label{fulld}
  \alpha (v,J,M)=\alpha^{e}(v,J,M)+\alpha^{v}(v,J,M)+\alpha^{r}(v,J,M)
\end{equation}
where the superscripts $e,v,r$ imply the electronic, vibrational
and rotational parts of the static dipole polarizability,
respectively, and the electronic part--$\alpha^{e}(v,J,M)$ can be
calculated by averaging over the mean electronic polarizability
(details of Eq.~(\ref{fulld}) can be found in \cite{bish4}).

Seeing Eq.~(\ref{fulld}), the rovibronic effects can be taken into
account by averaging over the vibrational and rotational wave
functions, of quantities calculated in the Born-Oppenheimer
approximation.

In the calculation of the mean electronic polarizability in this
\textit{Letter} the effects the electron mass respect to the
nuclei mass has not been taken into account. This effects,
however, must be taken into account when full three-body
Hamiltonian (details cane be found \cite{sher,tayl,moss2}) is
employed.

Using the unperturbed $\Psi_{0}$ and perturbed $\Psi'$ functions
obtained in this paper we can calculate some expectation values of
the lowest order relativistic correction (Eq.~(9) in \cite{kor4})
to the mean electronic polarizability $\alpha$ of the $1S\sigma$
state of $\mbox{H}^{+}_{2}$ molecular ion. In addition, the
accurate data used in evaluation of the mean electronic
polarizability $\alpha $ in this work can be employed in the
calculation of relativistic correction to it. This work is in
progress now.

\appendix
\section*{Appendix: Analytical evaluation of the matrix elements}
\renewcommand{\theequation}{A-\arabic{equation}}
\setcounter{equation}{0}

The calculation of the matrix elements is reduced to evaluation of
integrals of the type
\begin{equation}
 \Gamma_{lm}(\alpha,\beta) =
   \int r^{l-1}_{1}r^{m-1}_{2}e^{-\alpha r_{1}-\beta r_{2}}
                                            d^{3}\mathbf{r}.
\end{equation}
Integers $(l, m)$ are, in general, non-negative, but in case of singular
matrix elements one of the indices can be negative.

The function $\Gamma_{00}$ can be easily obtained
\begin{equation}\label{A.2}
\Gamma_{00}(\alpha,\beta,R) = \frac{4\pi}{R}\,
    \frac{e^{-\beta R} - e^{-\alpha R}}{\alpha^{2} - \beta^{2}},
\end{equation}
where $R$ is the distance between nuclei, then
$\Gamma_{lm}(\alpha,\beta;R)$ for non-negative $(l, m)$ may be generated
from (\ref{A.2}) by means of relation
\begin{equation}\label{A.3}
\Gamma_{lm}(\alpha,\beta;R) =
  \left( - \frac{\partial}{\partial \alpha}\right)^{l}
  \left( - \frac{\partial}{\partial \beta}\right)^{m}
  \Gamma_{00}(\alpha,\beta,R).
\end{equation}

Integral $\Gamma_{-1,0}(\alpha,\beta;R)$ is expressed by
\begin{equation}\label{A.4}
\begin{array}{@{}l}
\displaystyle
\Gamma_{-1,0}(\alpha,\beta;R) =
   \frac{2\pi}{R\beta}
      \Bigl\{
         e^{\beta R} \mbox{E}_{1}(R(\alpha+\beta))
         +e^{-\beta R}\ln R(\alpha +\beta)
\\[3mm]\hspace{30mm}
         -e^{\beta R}\bigl[\mbox{E}_{1}(R(\alpha -\beta))
         +\ln R(\alpha -\beta)\bigr]
      \Bigr\}.
\end{array}
\end{equation}

Worthy to note that a function in square brackets is analytic when
argument is zero. Integrals $\Gamma_{-1,m}$ are generated from
$\Gamma_{-1,0}$ similar to (\ref{A.3}):
\begin{equation}\label{A.5}
\Gamma_{-1,m}(\alpha,\beta;R) =
  \left( - \frac{\partial}{\partial \beta}\right)^{m}
  \Gamma_{-1,0}(\alpha,\beta,R).
\end{equation}

Function $\mbox{E}_1(z)$ encountered in (\ref{A.4}) is the exponential
integral function \cite{Abr}:
\[
 \mbox{E}_{1}(z) = \Gamma(0,z) = \int^{\infty}_{z} t^{-1}e^{-t}dt.
\]

\end{document}